\documentclass[11pt]{article}
\usepackage{graphicx}
\usepackage{epstopdf}
\usepackage{authblk}
\begin{document}

\title{Comparing open and closed molecular self-assembly}

\author{M. Castelnovo, T. Verdier}
\affil{Laboratoire de Physique, Ecole Normale Sup\'erieure de Lyon, Universit\'e de Lyon and CNRS, 46 All\'ee d'Italie, 69364 LYON Cedex 07, FRANCE}
\author{L. Foret}
\affil{Laboratoire de Physique Statistique, Ecole Normale Sup\'erieure, Universit\'e Pierre et Marie Curie, Universit\'e Paris Diderot and CNRS, 24 Rue Lhomond, 75005 PARIS, FRANCE}
\maketitle

\abstract{
We study theoretically in the present work the self-assembly of molecules in an open system, which is fed by monomers and depleted in partial or complete clusters. Such a scenario is likely to occur for example in the context of viral self-assembly. We provide a general formula for the mean-field size distribution which is valid both at equilibrium in a closed system, and in the stationary state in an open system. This allows us to explore in a simple way out-of-equilibrium features for self-assembly and compare them to equilibrium properties. In particular, we identify a region of parameter space for which the out-of-equilibrium size distribution in the presence of external fluxes is equal to the equilibrium size distribution in the absence of external fluxes, up to a constant renormalization factor. The range of validity of this result and its consequences are discussed.}

\section{Introduction}

Molecular self-assembly is a simple and yet powerful process allowing single molecules to reach collective properties spontaneously thanks to favorable interactions. It is widely observed in nature in order to organize molecules. The most significant example is the self-assembly of lipid or lipid-like molecules into the various membranes of every cell \cite{lodish}. The quantitative description of self-assembly is classically performed within the framework of micellization thermodynamics. This approach is well-suited in order to predict for example the appearance of a \textit{critical micellar concentration (CMC)}, or the cluster equilibrium size distribution \cite{gelbartBOOK}. Historically, this framework was introduced in order to describe \textit{in vitro} experiments in which a given amount of molecules is solubilized, leading to clustering of the molecules. However, for many cases in biology, self-assembly is an \textit{open} process in which the monomer units are constantly produced and the final configurations of clusters are frozen or removed from the monomer-cluster exchange process. Examples of such open self-assembly systems are cell membranes with inhomogeneous composition \cite{turner2005,dmitrieff2011}, or viral self-assembly of envelopped or non-envelopped  viruses \cite{zlotnick1994,zlotnick1999,zlotnick2007}. In the latter cases of viral self-assembly, the monomer source is associated to production of capsid protein within a well defined time frame. These particular proteins tends to self-assemble in order to build a thin shell aiming at protecting its genome. In the case of envelopped viruses, the self-assembly takes place at one of the cell membrane, and its final product is released out of the cells, while for non-envelopped viruses, the self-assembly takes place in cells, and the final product is a closed protein shell. In both cases, the final products of self-assembly are not able to exchange monomers with incomplete shells with intermediate completion. For all these open systems, equilibrium considerations are strictly not expected to be applicable, although they are widely used in order to perform zeroth order modeling of self-assembly \cite{vanderschoot2005,zandi2009,castelnovo2013}. 

The aim of this work is to study theoretically the expected deviations between clusters populations in the absence or the presence of external material exchanges. In order to extend the classical equilibrium description of closed self-assembly to an open one, we use the mean-field kinetic equations describing the open self-assembly process, allowing for constant monomer input  and cluster removal. In the case of a closed system, these equations are known as the Becker-D\"oring (BD) equations \cite{becker1935}, and their rich behavior has been extensively studied \cite{ball1986,coveney1998,penrose1997,wattis2004,wattis2006,dorsogna2012,yvinec2012}. The asymptotic state for these equations corresponds to the regular equilibrium description of self-assembly as provided by mass action and mass conservation laws. In the particular context viral self-assembly, some out-of-equilibrium features have been predicted, associated to the particular energy landscape in this case \cite{zandi2006,morozov2009}. In the case of open systems, results are more scarce. The main results have been obtained on BD equations on infinite systems in the presence of monomer input. In this case, self-similar clusters populations have been shown to arise \cite{wattis2004}. Within the context of biological-oriented applications, several authors showed also for example the existence of finite size clusters on membrane under continuous recycling scheme, with sizes compatible with typical lipid rafts \cite{turner2005,foret2012}. Similarly, Foret and Sens uncovered a realistic mechanism of kinetic regulation of coated vesicle secretion based on Becker-D\"oring type of modeling of open self-assembly \cite{foret2008}.

Within this work we use the general framework of mean-field kinetic equations of molecular self-assembly in order to identify the characteristic traits of open self-assembly models, as compared to the equivalent closed self-assembly models. In the particular context of biological self-assembly, the present approach allows to investigate the following questions: \textit{(i)} is it possible to use the informations about cluster energetics and external exchange rates (input and output) in order to predict exactly the stationary size distribution, and \textit{(ii)} how these distributions are compared with the one of equivalent closed system in the absence of external fluxes?
Our analysis of the kinetic equations in the stationary limit allows to identify the relevant combination of variables for a finite system in order to answer these two questions. In the general case of a single input and arbitrary distribution of outputs, we provide recursive relations allowing to compute exactly the stationary cluster size distribution. In the particular case of a single monomer input and a single cluster output for the maximal cluster size, we found an exact solution for the stationary size distribution of clusters as function of internal and external exchange rates.  Moreover, our analysis shows that the cluster size distribution in this case is well approximated by a renormalized equilibrium distribution, with a factor that is weakly size dependent. The range of validity of this constant renormalization regime is discussed.

\section{Results}
\subsection{General case} 
\begin{figure*}[htbp]
\begin{center}
\includegraphics[scale=0.5]{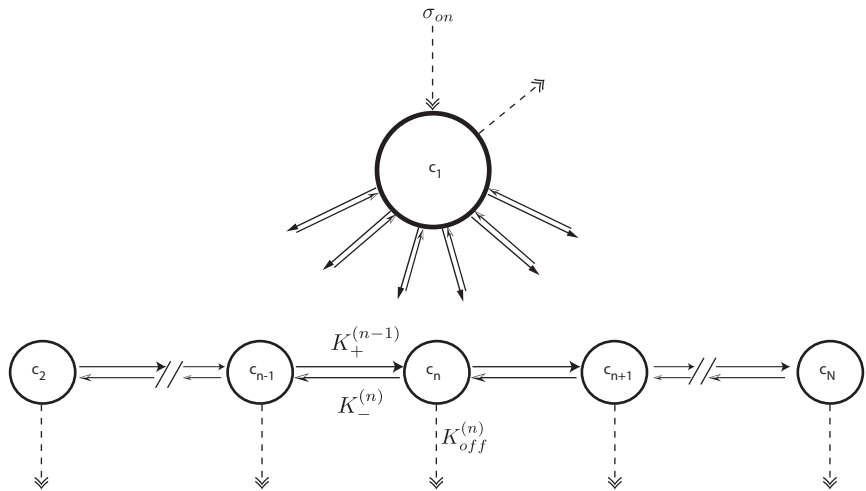}
\caption{Pathway of open self-assembly. The population of monomers $c_1$ is coupled to all others populations $\{c_n\}$, as it is indicated with double arrows. The dashed arrows are absent for closed self-assembly, while they are present for open self-assembly.}
\label{cartoon}
\end{center}
\end{figure*}

We consider molecules that tend to form clusters of variable size $n$ thanks to the repeated addition or removal of a single molecule. For the sake of simplicity, we impose an upper bound $N$ to the number of molecules in a cluster. This assumption is well-suited in the case of viral self-assembly, for which clusters cannot grow indefinitely (see figure \ref{cartoon}). The number of clusters of size $n$ is denoted $c_n$. These $n-$clusters can grow, shrink or escape the self-assembly process with respective rates $k_+^{(n)}$, $k_-^{(n)}$ and $k_{off}^{(n)}$. The energy of a single cluster of size $n$ is noted $E_n$. The relation between cluster energetics and internal rates is given by the standard detailed balance $k_+^{(n-1)}/k_-^{(n)}= e^{-\beta\left[E_n-E_{n-1}-E_1\right]}$, where  $\beta=1/kT$ is the inverse thermal energy. For the sake of simplicity, we assume  that $k_-^{(n)}=k_-$ for all $n$. Using the elementary rates, one can define respectively the forward, backward or off-fluxes at size $n$ by $K_+^{(n)}=k_+^{(n)}c_1c_n$, $K_-^{(n)}=k_-^{(n)}c_n$, $K_{off}^{(n)}=k_{off}^{(n)}c_n$. The net forward flux between size $n-1$ and $n$ is similarly defined as $J_n=K_{+}^{(n-1)}-K_-^{(n)}$. Taking into account the fact that the monomer population $c_1$ of clusters is fed with a constant rate $\sigma_{on}$, the mean-field kinetic equations \footnote{Note that these equations are slightly different from the one that Foret and Sens used to describe the kinetic regulation of coated vesicle secretion \cite{foret2008}. The main difference is that the escape rate in their case corresponds to the escape of \textit{monomers} from the self-assembly process on each particle, while we consider the escape of \textit{clusters}.} are written in a compact form in term of fluxes $J_n$ as 
\begin{eqnarray}
\dot{c}_1 & = & \sigma_{on}-K_{off}^{(1)}-J_2-\sum\limits_{n=2}^{N}J_n\label{monomerEq}\\
\dot{c}_n & = & -K_{off}^{(n)}+J_n-J_{n+1}\label{AllEq} \\
\dot{c}_N & = & -K_{off}^{(N)}+J_N \label{NEq}
\end{eqnarray}

Since the cluster population $c_n$ is strongly coupled to the evolution of monomer population $c_1$, these equations are non-linear, and a general time-dependent analytical solution is out of reach. Rather, useful analytical informations can be obtained by considering the stationary limit $\{\dot{c}_n\}=0$ for the system. The general solution of the stationary equations is obtained by using the \textit{ansatz} $c_n=A_nc_{n-1}$ for $n=2$ to $N$. Accordingly, the parameters $A_n$ obey the following recursive relations
\begin{eqnarray}
A_N & = & \frac{k_{+}^{(N-1)}c_1}{k_-^{(N)}+k_{off}^{(N)}} \label{sol1}\\
A_n & = & \frac{k_{+}^{(n-1)}c_1}{k_-^{(n)}+k_{off}^{(n)}+k_+^{(n)}c_1-k_-^{(n+1)}A_{n+1}} \, \,\, \,  \label{sol2}
\end{eqnarray}
where the last relation is valid for $n=2,...N-1$. These general formula depend on the various rates $\{k_+^{(n)},k_-^{(n)},\sigma_{on},k_{off}^{(n)}\}$ and on the monomer population $c_1$ in the stationary state. The complete stationary solution is obtained by using in backward way the recursive relations Eq.\ref{sol1} to \ref{sol2} into the stationary balance for monomers (Eq.\ref{monomerEq} with $\dot{c}_1=0$) and by solving it for the only remaining unknown $c_1$. The present recursive relations allows therefore to compute \textit{exactly} in a simple way the cluster populations $c_n=\left(\prod\limits_{g=2}^{g=n}A_g\right)c_1$ both in  closed and open systems, once all the internal and external exchange rates are provided.

A careful analysis of the previous recursive equations allows to identify the main differences between a system in the absence and in the presence of external rates quantified by the monomeric input $\sigma_{on}$ and cluster outputs $k_{off}^{(n)}$. Indeed, in the limit of closed self-assembly, for which all the external rates vanish, the recursive relations are decoupled and the solution is given by $A_n^{(eq)}=\frac{k_+^{(n-1)}c_1}{k_-^{(n)}}$. One recovers this way the detailed balance of kinetic equations or equivalently the classical mass action law $c_n^{(eq)}=\frac{k_+^{(n-1)}}{k_-^{(n)}}c_1c_{n-1}^{(eq)}$, as expected. The iteration of these relations together with the use cluster energetics $E_n$ gives the standard form for the cluster size distribution at equilibrium $c_n^{(eq)}=c_1^ne^{-\beta (E_n-nE_1)}$. 

In the presence of inputs and outputs, the introduction of a new parameter $H_n$ defined by the relation $A_n=\frac{k_+^{(n-1)}c_1}{k_-^{(n)}}\left(1-H_n\right)$ allows to show that the cluster size distribution can always be written as a renormalized equilibrium cluster distribution through the relation $c_n=c_n^{(eq)}R_n$ with $R_n=\prod\limits_{g=2}^{g=n}\left(1-H_n\right)$. Note that the renormalization factor $R_n$ is generally \textit{size}-dependent. The quantity $H_n$ is interpreted as a measure of local deviation from equilibrium due to the presence of external rates. This can be understood from the following exact relations derived from the stationary equations:
\begin{equation}
\frac{K_-^{(n)}}{K_+^{(n-1)}}=1-H_n \label{Hndef} \\
\end{equation}
In the absence of external fluxes, the ratio between backward and forward fluxes is equal to unity at every links of the chain of reactions, equivalently to the detailed balance condition. Local deviations from equilibrium are therefore quantified by the parameter $H_n$, as it is suggested by Eq.\ref{Hndef}.

With the aim of discussing first the consequence of formula Eq.\ref{sol1} and \ref{sol2}, we introduce in this work a toy model for cluster energetics. We assume a simple quadratic energy dependence on the size $E_n=E_0((n-n_0)^2-(1-n_0)^2)$, giving rise, at equilibrium, to a cluster size distribution with a single peak. Using this choice, we first compute the modulation of cluster population by the presence of a single monomeric source and multiple outputs for different sizes. We changed both the values and localizations of cluster escape rates. In particular, we assume a step function for the escape rate $k_{off}^{(n)}$: it vanishes for $n<n_*$, and takes a finite value $k_{off}$ for $n>n_*$. This choice mimicks the presence of a size threshold for clusters removal from self-assembly.
\begin{figure}[htbp]
\begin{center}
\includegraphics[height=7cm]{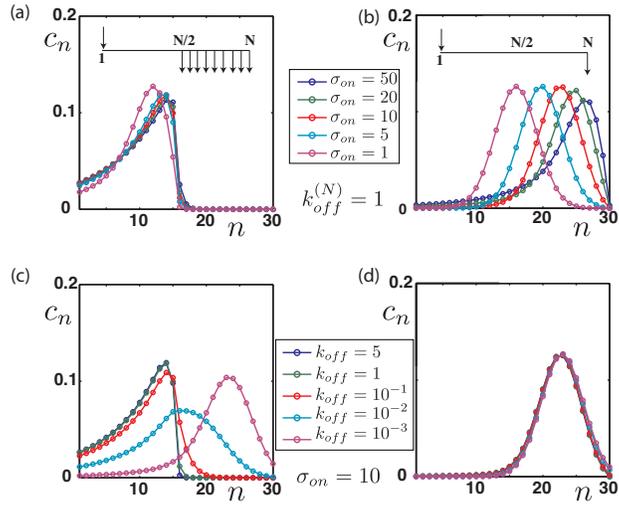}
\caption{Stationary size distribution for open self-assembly, as function of localization and amplitude of escape rates. Size distributions at fixed escape rate ($k_{off}^{(N)}=1$) for various monomeric fluxes are shown for extended outputs \textit{(a)} and localized output \textit{(b)}. Size distributions at fixed monomeric input ($\sigma_{on}=10$) for various outputs are shown for extended outputs \textit{(c)} and localized output \textit{(d)}. Parameters used to compute this figure: $N=30,kT=1,k_-^{(n)}=0.05,E_0/kT=0.05,n_0=4N/5$.}
\label{figure2}
\end{center}
\end{figure}
The variations of size distribution associated to the change of both the escape rate values $k_{off}$ and source term $\sigma_{on}$ is shown in figure \ref{figure2}. These changes are larger for strong monomer source than for weak monomer source. In the case of unique localization of the escape rate to the maximal size of the system, the modulation of size distribution is also observed, although with a weaker amplitude. 

\subsection{One input and one output} 
\begin{figure}[htbp]
\begin{center}
\includegraphics[height=10cm]{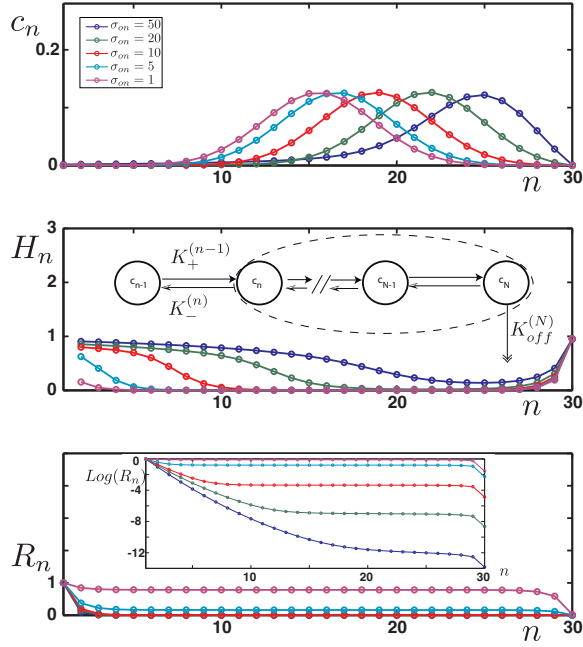}
\caption{Stationary cluster populations, and local deviation from equilibrium. \textit{(a)} Cluster population for different monomer input at fixed cluster output $k_{off}^{(N)}=1$. \textit{(b)} Local deviation computed according to Eq.\ref{Hn_exact} for the population of $(a)$. The inset provides natural interpretation for local deviations defined as $H_n=K_{off}^{(N)}/K_{+}^{(n-1)}$. \textit{(c)}. Renormalization factors $R_n$. The inset shows the logarithm of $R_n$, in order to highlight constant $R_n$. Parameters used to compute this figure:$N=30,kT=1,k_-^{(n)}=0.05,E_0/kT=0.05,n_0=N/2$.}
\label{figurelocalmode}
\end{center}
\end{figure}

In the particular case of one monomer source and a single cluster output for the maximal size $N$, further \textit{exact} analytical results can be obtained. First, close inspection of stationary equations shows that the net flux $J_n=K_{+}^{(n-1)}-K_-^{(n)}$ is independent of the size $n$. The value of this common flux $J$ is such that $J=\sigma_{on}/N=K_{off}^{(N)}$, and it vanishes in the absence of external fluxes. The balance between backward and forward flux is however size-dependent, and it is quantified by $H_n$ through Eq.\ref{Hndef}. It can also be shown that $H_n=K_{off}^{(N)}/K_{+}^{(n-1)}$. This relation tells us that the presence of a cluster output for maximal size induces an imbalance between backward and forward fluxes that propagates from the maximal size cluster down to the cluster of size $n$. Furthermore, after some algebra, the local deviation from detailed balanced $H_n$ defined above is written exactly as:
\begin{equation}
H_n=\frac{\left(c_N^{(eq)}/c_n^{(eq)}\right)F_n}{1+\sum\limits_{g=n}^{g=N-1}\left(c_N^{(eq)}/c_g^{(eq)}\right)F_g } \,\,\, \mathrm{for }\,\,\,n=2,...N-1
\label{Hn_exact}
\end{equation}
with $F_n=\frac{k_{off}^{(N)}/k_-^{(n)}}{1+(k_{off}^{(N)}/k_-^{(N)})}$ and $H_N=F_N$. With our choice of constant $k_-^{(n)}$, the parameters $F_n$ are independent of the size. The ratio of equilibrium size distribution are rewritten as $c_N^{(eq)}/c_n^{(eq)}=e^{-\beta\Delta G_{n\rightarrow N}}$ with $\Delta G_{n\rightarrow N}=E_N-E_n-(N-n)(E_1+\ln c_1)$. This free energy difference is associated to the cost for growing a cluster of size $n$ to the maximal size $N$.
In order to illustrate these deviations, we use the same quadratic cluster energetics as in Figure \ref{figure2}. The size distribution $c_n$ and the local deviation parameters $H_n$ are presented in figure \ref{figurelocalmode} for different monomer inputs. 
The main observation about these local deviations is that they are not uniform. Interestingly, the deviations are the smallest within a range around the maximal cluster population. This means that in this region, the self-assembly reactions occur locally very close to equilibrium, and the detailed balance is partially obeyed, while significant deviations from equilibrium are restricted outside this region.

Using the previous formula for local deviation $H_n$, or equivalently the presence of a constant flux $J$ throughout the system, the size distribution is cast in the form $c_n=c_n^{(eq)}R_n$, with the renormalization factors $R_n$ defined by:

\begin{eqnarray}
R_n & = & \frac{1+\sum\limits_{g=n+1}^{g=N-1}\left(c_N^{(eq)}/c_g^{(eq)}\right)F_g}{1+\sum\limits_{g=2}^{g=N-1}\left(c_N^{(eq)}/c_g^{(eq)}\right)F_g} \, \, \mathrm{for}\,n<N-1\nonumber\\
R_{N-1} & = & \frac{1}{1+\sum\limits_{g=2}^{g=N-1}\left(c_N^{(eq)}/c_g^{(eq)}\right)F_g} \\
R_{N} & = & \frac{1-F_N}{1+\sum\limits_{g=2}^{g=N-1}\left(c_N^{(eq)}/c_g^{(eq)}\right)F_g }\nonumber
\end{eqnarray}
The values of these renormalization factors are shown in figure \ref{figurelocalmode} for various monomeric inputs. Interestingly, these factors are roughly independent of the size $n$ over a large range of parameters. This observation is in agreement with the weak values of parameters $H_n$ mentioned previously. As a consequence, the size distribution in the presence of localized external input and outputs can be roughly described by the size distribution in their absence, up to a constant renormalization factor smaller than unity.
 
\begin{figure}[htbp]
\begin{center}
\includegraphics[height=13cm]{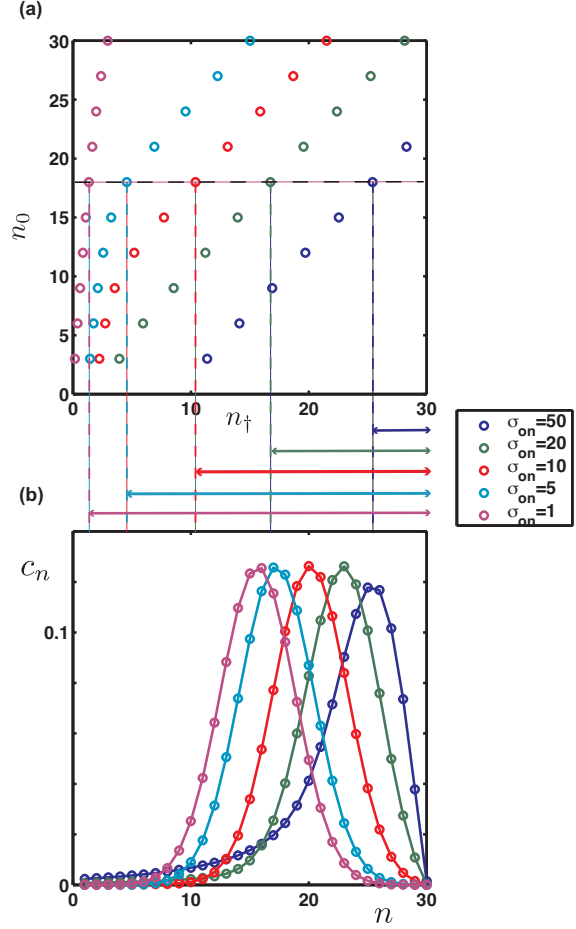}
\caption{Range of validity of constant renormalization regime $[n_{\dagger},N]$. \textit{(a)} The position of $n_{\dagger}$ is shown as function of both $n_0$ and $\sigma_{on}$. \textit{(b)} Size distributions as function of $\sigma_{on}$ for the particular value $n_0=3N/5$  highlighted in \textit{(a)} by the horizontal black dashed line. The vertical colored dashed lines show the value of $n_{\dagger}$. The horizontal double arrows represent the range of sizes where the constant renormalization regime is expected to be valid. Each colors correspond to the legend on the right of the figure.Parameters used to compute this figure:$N=30,kT=1,k_-^{(n)}=0.05,E_0/kT=0.05$.}
\label{Ndag}
\end{center}
\end{figure}

The functional form for renormalization factor allows to find out the range of validity of the constant renormalization regime. Indeed, by going to the continuous limit for these equations, it is found that the slope of renormalization factor scales like $R'_n\sim e^{-\beta \Delta G_{n\rightarrow N}}$. Therefore the region of small slope associated to the constant renormalization regime is found approximately for positive values of free energy variations $\Delta G_{n\rightarrow N}>0$. Within the context of our energetic toy model for which $E_n=E_0((n-n_0)^2-(1-n_0)^2)$, this condition is written exactly as $n_\dagger < n <N$ with 
\begin{equation}
n_{\dagger}=2n_0-N+\frac{\ln c_1}{E_0}
\end{equation}
The width of the interval $N-n_{\dagger}$ depends therefore on the localization of energy minimum $n_0$, on the width of the energy around this minimum $E_0$ and on the value of monomer population $c_1$. As a rule of thumb, the constant renormalization regime is therefore expected to be observed if the energy minimum is localized far from the maximal size $N$ (small values of $n_0$), and if the input strength measured by $\sigma_{on}$ is moderate (small values of $c_1$). If these conditions are not met, the size distribution cannot be approximated by using the equilibrium distribution.  The dependence in both $n_0$ and $c_1$ is illustrated in figure \ref{Ndag}.
For a given choice of $n_0$, it is indeed observed that the range of validity of constant renormalization regime is much larger than width of the distribution at low $\sigma_{on}$, while this range is very small at large $\sigma_{on}$.

\section{Discussion}

In this work, we investigated the influence of localized inputs and outputs on the size distribution of cluster self-assembly, as compared to the equivalent "closed" system with the same internal rates. We obtained general analytical formula for the size distribution in terms of recursive relations. We provide local measurements of deviation from detailed balance induced by the presence of external fluxes. Focusing on the special case of one single monomeric input and one single cluster output for maximal size $N$, we obtained further exact results. Our main result in this case is that, under some conditions, the out-of-equilibrium size distribution is well approximated by a renormalized equilibrium size distribution, with a factor that is weakly size-dependent. As a consequence, the size distribution follows the equilibrium distribution $c_n=c_1^ne^{-\beta (E_n-nE_1)}R_n$ with an appropriate monomer population. In other words, localized external fluxes perturb only moderately the cluster growth, and change essentially the effective monomer population. Moreover, we showed that both the renormalization factor $R_n$ and the local deviation parameter $H_n$ depend critically on the free energy variations $\Delta G_{n\rightarrow N}$ necessary to grow a cluster of size $n$ to the maximal size $N$. Indeed, detailed balance within equilibrium self-assembly in a closed system is the result of a process in which clusters of size $n$ equilibrate their \textit{in-fluxes} from neighbouring sizes $n-1$ and $n+1$ with their \textit{out-fluxes} toward the same neighbouring sizes thanks to the presence of monomers. The presence of external inputs and outputs modifies this equilibration both globally and locally. We showed in particular the presence of a constant flux $J$ throughout the sizes of clusters. The influence of external fluxes at a local scale are understood by considering free energy variations. Indeed, if the free energy variation $\Delta G_{n\rightarrow N}$ is negative in some range, the growth of clusters is a downhill process from size $n$ to size $N$, and
large deviations to detailed balance due to the output of the largest clusters propagate along this range of sizes. On the contrary, for $\Delta G_{n\rightarrow N}>0$, the growth from $n$ to $N$ is unfavorable, and deviations from detailed balance caused by output flux at the maximal size weakly propagate down to size $n$. The latter condition is favorable to the observation of the constant renormalization regime. The identification of such a regime provide on the one hand a partial and natural justification for the use of equilibrium size distribution in order to perform zeroth-order modeling of molecular self-assembly in many biological systems, and on the other hand it provides a way to quantify the range of validity of such an approximation. Interestingly, it has also been shown by some authors that, although  \emph{closed} self-assembly is expected theoretically to show out-of-equilibrium behavior in some parameter range, as mentioned in the introduction\cite{morozov2009,zandi2006}, a \emph{pseudo}-law of mass action is obeyed at finite time. This shows again that apparent equilibrium description of self-assembly systems is a valid first-step approach within another context.

When multiple outputs are considered, like those shown in figure \ref{figure2}a and \ref{figure2}c, the influence of outputs is expected to be felt at larger scale. This additional complexity prevented us to find analytical results about local deviation parameter $H_n$ or renormalization factors $R_n$. Rather, numerical analysis based on the recursive relations Eq.\ref{sol1} to \ref{sol2} allow to compute numerically the size distribution once the energetics of cluster formation together with the external rates are provided. In this case, these general formula allows to go beyond the simple equilibrium modeling and to take into account the effect of external fluxes on the self-assembly process.

\emph{Acknowledgements} --
The authors would like to thank the \textit{Fondation Simone and Cino Del Duca} from Institut de France and CNRS (PEPS PTI 2013) for initial funding of this work.

\bibliographystyle{unsrt}
\bibliography{bib}

\end{document}